\documentclass{cimento}

\title{Soft behavior of string amplitudes with external massive states}
\author{A.~L.~Guerrieri\from{ins:x}}
\instlist{\inst{ins:x} Dipartimento di Fisica, Universit\`a di Roma ``Tor Vergata" and Sezione INFN di Roma II ``Tor Vergata"}

\PACSes{\PACSit{11.25}{w}}

\begin{document}

\maketitle

\begin{abstract}
We briefly discuss the soft behavior of scattering amplitudes both in string and quantum field theory. In particular we show a general argument about the validity of soft theorems for open superstring amplitudes and list some of our recent results. 
\end{abstract}

Soft behavior of scattering amplitudes in QED~\cite{Low:1954kd} and gravity~\cite{Weinberg:1965nx} has been studied long time ago. 
Recently it received renewed attention due to  its connection~\cite{Strominger:2013jfa, He:2014laa, Cachazo:2014fwa} with the BvBMS symmetry~\cite{Bondi:1962px}.
Supertring amplitudes of massless open and closed strings show the same universal soft behavior as the corresponding amplitudes in quantum field theory~\cite{Bianchi:2014gla}. This result is expected to hold if one observes that the three-linear coupling of massless open superstrings is the same as the Yang-Mills coupling of three gluons and the coupling of three gravitons in the Type II theory is the same as in Einstein gravity.
Moreover open bosonic string amplitudes behave universally even in presence of a correction $F^3$ to the Yang-Mills vertex and the coupling with the tachyon $\mathcal{T} F^2$, while closed bosonic string amplitudes don't due to the presence of a tree-level non minimal coupling $\phi R^2$ with the dilaton. Recently we pointed out that also amplitudes involving massive string excitations behave universally when the momentum of one of the massless legs becomes soft~\cite{Bianchi:2015yta}.
In~\cite{Bianchi:2014gla, Bianchi:2015yta} the soft limit of open string amplitudes was studied abstractly using the OPE analysis of string vertex operators. The conclusion was that string amplitudes of gluons and higher-spin massive excitations behave universally, in formulae
\begin{equation}
\mathcal{A}_n(1.{.}.s.{.}.n)=\left\{ \left[\frac{a_sk_{s+1}}{k_sk_{s+1}}-\frac{a_sk_{s-1}}{k_sk_{s-1}}\right] + \left[ \frac{f_s {:}J_{s+1}}{k_sk_{s+1}} -\frac{f_s{:}J_{s-1}}{k_sk_{s-1}}\right] \right\}\mathcal{A}_{n-1}(1.{.}.\hat{s}.{.}.n),
\label{theo}
\end{equation}
with $J_i$ the angular momentum operator of the $i$-th external leg.
The proof follows if one considers the OPE of the soft massless leg with the adjacent vertex operators. More concretely, let's consider the case in which both the adjacent operators are gluons and without loss of generality suppose that all vertices are integrated. The OPE gives
\begin{equation}
\int^{z_{s+1}} dz_s~(z_{s+1}-z_s)^{2\alpha^\prime k_sk_{s+1}-1}F(z_s,z_i)\approx \frac{F(z_{s+1},z_i)}{2\alpha^\prime k_sk_{s+1}}
\end{equation}
when $k_s\to 0$ and similarly for the operator in $z_{s-1}$. At the leading order in the soft momentum $k_s$ one gets
\begin{equation}
V_A(k_s,z_s)V_A(k_{s\pm 1},z_{s\pm 1})\approx \frac{a_sk_{s\pm 1}}{k_sk_{s\pm 1}}V_A(a_{s\pm 1},k_{s\pm1}+k_s).
\end{equation}
The inclusion of higher order corrections to the OPE and vertex operators for higher-spin states allows one to reproduce the statement contained in eq.~\ref{theo}. 
We computed explicitly 4-point superstring amplitudes with 3 gluons and one state belonging to the first massive level. In particular, we considered the symmetric, transverse and traceless $H$ and the totally antisymmetric and transverse $C$ state represented respectively in the $-1$ super-ghost picture by the vertex operators
\begin{equation}
V_H=H_{\mu\nu}i\partial X^\mu \psi^\nu e^{ipX},\quad V_C=C_{\mu\nu\rho}\psi^\mu\psi^\nu\psi^\rho e^{ipX},
\end{equation}
obtaining exactly the behavior predicted by the OPE analysis in eq.~\ref{theo}.
We also checked the validity of a recently proposed formula~\cite{Mafra:2011nv} expressing $n$-point amplitudes of massless open strings in terms of Yang-Mills field theory amplitudes.
Exploited the unitarity of string theory we obtained the amplitudes with massive states factorising on the massive pole in some 2-particle channel. Using dimensional reduction we also performed the same computation with 4-dimensional momenta and polarisations in the spinor helicity formalism~\cite{Elvang:2013cua}.
Therefore we propose to compute $n$-point amplitudes with $n$ massive legs factorising $2n$-point massless amplitudes, avoiding in this way the complication due to the OPE for higher-spin operators.
Using KLT relations it is possible to efficiently compute closed string amplitudes with massive states as the ``square" of the open ones we computed in~\cite{Bianchi:2015yta}.
We plan to investigate the soft behavior of tree-level closed scattering amplitudes confirming the universality or the lack of it in the Type IIA, Type IIB, bosonic and heterotic case.
We hope also to shed further light on the soft theorems for the Kalb-Ramond, the dilaton~\cite{DiVecchia:2015oba} and the other moduli fields.

%
%
%
%
%
%
%
%


\begin{thebibliography}{0}

\bibitem{Low:1954kd} 
  F.~E.~Low,
  Phys.\ Rev.\  {\bf 96}, 1428 (1954).

\bibitem{Weinberg:1965nx} 
  S.~Weinberg,
  Phys.\ Rev.\  {\bf 140}, B516 (1965).
  D.~J.~Gross and R.~Jackiw,
  Phys.\ Rev.\  {\bf 166}, 1287 (1968).

\bibitem{Strominger:2013jfa} 
  A.~Strominger,
  JHEP {\bf 1407}, 152 (2014)
  [arXiv:1312.2229 [hep{-}th]].

\bibitem{He:2014laa} 
  T.~He, V.~Lysov, P.~Mitra and A.~Strominger,
  arXiv:1401.7026 [hep-th].

\bibitem{Cachazo:2014fwa} 
  F.~Cachazo and A.~Strominger,
  arXiv:1404.4091 [hep-th].

\bibitem{Bondi:1962px} 
  H.~Bondi, M.~G.~J.~van der Burg and A.~W.~K.~Metzner,
  Proc.\ Roy.\ Soc.\ Lond.\ A {\bf 269}, 21 (1962).

\bibitem{Bianchi:2014gla} 
  M.~Bianchi, S.~He, Y.~t.~Huang and C.~Wen,
  arXiv:1406.5155 [hep-th].

\bibitem{Bianchi:2015yta} 
  M.~Bianchi and A.~L.~Guerrieri,
  arXiv:1505.05854 [hep-th].

\bibitem{Mafra:2011nv}
  C.~R.~Mafra, O.~Schlotterer and S.~Stieberger,
  Nucl.\ Phys.\ B {\bf 873} (2013) 419
  [arXiv:1106.2645 [hep-th]].
    C.~R.~Mafra, O.~Schlotterer and S.~Stieberger,
  Nucl.\ Phys.\ B {\bf 873}, 461 (2013)
  [arXiv:1106.2646 [hep-th]].

\bibitem{Elvang:2013cua} 
  H.~Elvang and Y.~t.~Huang,
  arXiv:1308.1697 [hep-th].

\bibitem{DiVecchia:2015oba} 
  P.~Di Vecchia, R.~Marotta and M.~Mojaza,
  arXiv:1502.05258 [hep-th].



\end{thebibliography}
\end{document}